\documentclass[a4paper,nofootinbib,preprintnumbers,twocolumn,preprintnumbers,floatfix,superscriptaddress,pra,showpacs]{revtex4-1}

\usepackage{graphicx}
\usepackage{amssymb}
\usepackage{amsmath}
\usepackage{epstopdf}
\usepackage{graphics}
\usepackage[caption=false]{subfig}
\usepackage{float}
\usepackage{color}
\usepackage{hyperref}

\allowdisplaybreaks[1]

\newcommand{\ket}[1]{|#1\rangle}					

\newcommand{\figref}[1]{Fig.~\ref{#1}}

\begin{document}

\title{Efficient quantum computation in a network with probabilistic gates and logical encoding}

\author{J. Borregaard }
\affiliation{Department of Physics, Harvard University, Cambridge, MA 02138, USA}
\author{A. S. S\o rensen}
\affiliation{The Niels Bohr Institute, University of Copenhagen, Blegdamsvej 17, DK-2100 Copenhagen \O, Denmark}
\author{J. I. Cirac}
\affiliation{Max-Planck Institut f\"{u}r Quantenoptik, Hans-Kopfermann-Str. 1, D-85748 Garching, Germany}
\author{M. D. Lukin}
\affiliation{Department of Physics, Harvard University, Cambridge, MA 02138, USA}

\date{\today}

\begin{abstract}
A new approach to efficient quantum computation with probabilistic gates is proposed and analyzed in both a local and non-local setting. It combines heralded gates previously studied for atom or atom-like qubits with logical encoding from linear optical quantum computation in order to perform high fidelity quantum gates across a quantum network. The error-detecting properties of the heralded operations ensure high fidelity while the encoding makes it possible to correct for failed attempts such that deterministic and high-quality gates can be achieved. Importantly, this is robust to photon loss, which is typically the main obstacle to photonic based quantum information processing. Overall this approach opens a novel path towards quantum networks with atomic nodes and photonic links.
\end{abstract}

\pacs{}

\maketitle
Quantum systems have the potential to revolutionize information technology by employing quantum computers~\cite{ladd1,nielsen1} and quantum cryptography~\cite{gisin1}. In particular, quantum networks~\cite{monroenetwork,Kimble} may enable novel applications ranging from distributed quantum computing~\cite{eisert1,beals,nickerson} and secure multipartite function evaluation~\cite{benor,dupuis} over cryptographic conferencing~\cite{bose1} to ultra-sensitive sensors~\cite{komar1}.  However, a major obstacle against distributing quantum information is the detrimental effect of transmission losses. A so-called quantum repeater was first presented as a solution to this problem in the context of two-party quantum communication~\cite{briegel,duan4}. In this approach, high-quality non-local entanglement is first created between quantum memories in a heralded fashion such that transmission losses only result in a finite success probability. Once the scheme is successful, the created high quality entangled states enable long distance communications through teleporation and deterministic local gate operations. The efficient light-matter coupling necessary for the entanglement generation can naturally be used to mediate the gate operation but such approaches suffer from photon losses and inefficiencies, which limit the gate fidelity. To overcome this, the concept of heralding was recently extended to gate operations~\cite{borregaard1,sahand,das,duankimble,ivan}. In these schemes, the quantum gates are heralded by a measurement outcome that distinguishes whether a photon loss has occurred or not, similar to the non-local entanglement generation schemes. As a result, high fidelity gate operations are obtained in a successful event while the involved qubits must be discarded if the gate fails.  

While heralded operations are directly applicable in quantum communication tasks~\cite{borregaard2}, their application in quantum computation is less straightforward. It has been demonstrated how fault-tolerant, one-way quantum computation is possible with probabilistic operations~\cite{li1,fujii,duan2} and in linear optical quantum computation (LOQC), the probabilistic operations are compensated by encoding in multi-photon states~\cite{knill,kok,gilchrist1,fabian}. Both these approaches are, however, associated with substantial resource overhead due to the necessity of generating large cluster states for one way quantum computing~\cite{raussendorf} and the upper bound of 50\% for the success probability of Bell measurements with linear optics~\cite{ying}. We present an alternative approach to universal quantum computation with probabilistic gates by combining heralded local operations previously studied for atomic qubits with logical encodings from LOQC. The heralding enables high fidelity operations while the logical encoding makes it possible to correct for failed attempts resulting in deterministic logical operations. Due to the higher success probability of the atomic gates, this approach is much less resource demanding than LOQC. Furthermore, the logical encoding enables computation within a circuit model, which circumvents the substantial qubit overhead of cluster state computation. 

The proposed procedure is ideally suited for networks based on optically connected atoms or atom-like systems. The long coherence time of atomic qubits makes them ideal candidates for memory nodes in a quantum network. Entanglement can be efficiently distributed over long distances with photonic links, which enables non-local gates through teleportation.~\cite{eisert1,anders3}. By combining local heralded operations on atomic qubits with photonic links and encoding, we thus exploit the advantages of both approaches to perform efficient quantum computation in a network.    

To illustrate the main idea, we consider combining the heralded controlled phase gate (CZ-gate) described in Ref.~\cite{borregaard1} with the parity encoding introduced in Refs.~\cite{gilchrist1,gilchrist2}. The heralded CZ gate of Ref.~\cite{borregaard1} couples qubits through a cavity field and is heralded by a measurement on an auxiliary atom. This gate has a success probability, approaching unity in the limit of strong emitter cavity coupling with a unity fidelity $F=1$ regardless of the coupling. A failed attempt of the gate leaks information about the two participating qubits to the environment corresponding to projecting onto the qubit basis. This type of error can be corrected (as described below) if the cavity couples to closed transitions in the qubits such that the interaction with the environment is an effective dephasing process. Similar situations can be encountered in other heralded gate schemes~\cite{sahand,das,duankimble,ivan} but we focus on the scheme of Ref.~\cite{borregaard1} for concreteness. 

The logic encoding of Refs.~\cite{gilchrist1,gilchrist2} was developed to protect against photon loss but as we show here, it can also be used to correct for the dephasing error following a failed attempt of the gate. The logical qubit states are encoded across $n$ physical qubits in the form of GHZ states $\ket{0^{n}}=\frac{1}{\sqrt{2}}\left(\ket{+}^{\otimes n}+\ket{-}^{\otimes n}\right)$ and $\ket{1^{n}}=\frac{1}{\sqrt{2}}\left(\ket{+}^{\otimes n}-\ket{-}^{\otimes n}\right)$ where $\ket{\pm}=\frac{1}{\sqrt{2}}\left(\ket{0}\pm\ket{1}\right)$. We will refer to $\left\{\ket{+},\ket{-}\right\}$ as the rotated basis and $\left\{\ket{0},\ket{1}\right\}$ as the qubit basis. A logical encoded qubit $\ket{\Psi^{n}}=a\ket{0^{n}}+b\ket{1^{n}}$ is protected from heralded errors leading to dephasing of a physical qubit. Following a dephasing error, the relevant qubit is measured in the qubit basis. If the qubit is measured to be in state $\ket{0}$, the remaining qubits will be in the state $\ket{\Psi^{n-1}}$ while if it is measured to be in state $\ket{1}$, the remaining qubits will be in state $a\ket{1^{n-1}}+b\ket{0^{n-1}}$, which is identical to $\ket{\Psi^{n-1}}$ up to a single qubit rotation. The logical state can thus be recovered after dephasing.  Consequently, this encoding can be used to correct for a failed CZ-gate by measuring the state of the physical qubits and applying appropriate single qubit rotations. Every failed gate will remove one level of encoding from the logical states. 

The logical qubits can be grown efficiently using methods known from cluster state preparation~\cite{gilchrist1,duan2}. To do this, we fuse together a $(n)$-qubit GHZ state with a $(m)$-qubit GHZ state by making a CNOT gate between one qubit from each GHZ state. We then measure the target qubit in the rotated basis and apply appropriate single qubit rotations based on the measurement result. The logical encoding means that if the CNOT gate fails, a qubit is removed from each of the two GHZ states resulting in $(n-1)$-qubit and $(m-1)$-qubit GHZ states. However, if the fusion succeeds a $(n+m-1)$-qubit GHZ state is prepared. Having prepared the resource state $\ket{0^{n}}$, a logical qubit  $\ket{\Psi^{n}}=a\ket{0^{n}}+b\ket{1^{n}}$ is prepared by applying the operation $U_{\psi}=a\mathrm{I}+b\sigma_{x}$, on a single qubit belonging to $\ket{0^{n}}$. Here $\mathrm{I}$ is the identity and $\sigma_{x}$ is a Pauli $x$-rotation of the qubit. 

\begin{figure} [h]
\centering
\includegraphics[width=0.45\textwidth]{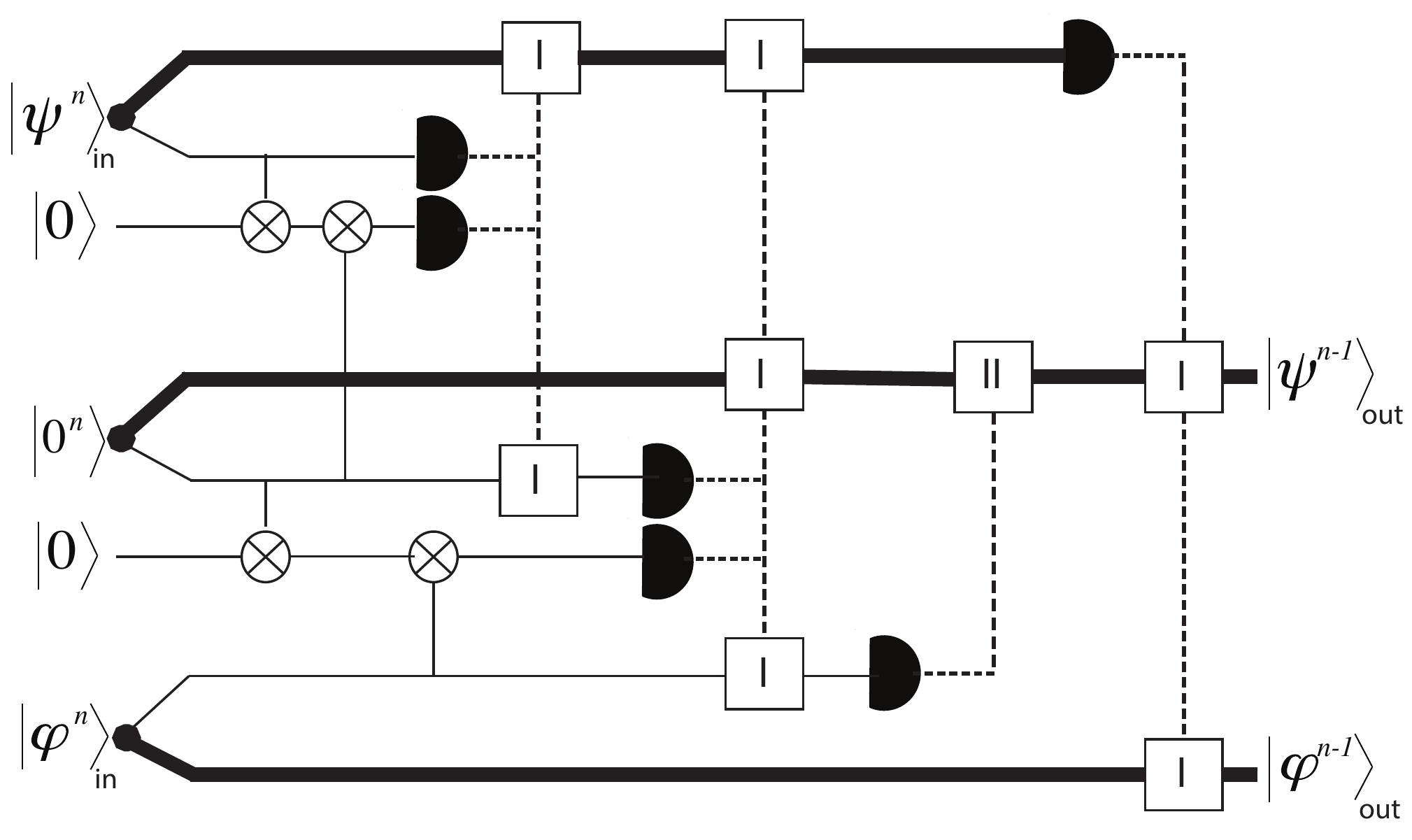}
\caption{Scheme of the logical CNOT gate. Crossed circles represent effective CNOT-gates with the cross marking the target qubit. Filled half-circles represent detectors, thick lines represent multi-qubit states and narrow lines are single qubit states. Based on the measurement results, Pauli rotations $\sigma_{z}$ or $\sigma_x$ are applied to either a single qubit (I) or multiple qubits (II). The final multi-qubit detection is a parity measurement. The scheme makes a logical CNOT between $\ket{\psi}$ (control) and $\ket{\phi}$ (target) and consumes at least one physical qubit from each state. Note that $\ket{\psi}$ is swapped to the resource state $\ket{0^n}$ in the process.  }
\label{fig:CNOTgate}
\end{figure}

A universal set of quantum gates within the logical encoding can be obtained through Bell measurements, single qubit rotations, and an auxiliary multi-qubit resource state ($\ket{0^n}$) as described in Ref.~\cite{gilchrist2}. The multi-qubit resource state is necessary for some of the gates in the universal set of which the logical CNOT gate is the most demanding~\cite{gilchrist1} and we therefore focus on this. We transform the logical CNOT of Ref.~\cite{gilchrist1} into our setup employing heralded CZ gates and auxiliary qubits. A schematic of the corresponding logical CNOT gate is shown in \figref{fig:CNOTgate}. Note that the logical CNOT, in principle, only requires two CNOT gates between physical qubits but we require two additional CNOT gates with auxiliary qubits in our setup. The auxiliary qubits are necessary because we are not making a direct CNOT gate between physical qubits but an effective CNOT gate using single qubit rotations and a CZ-gate. An effective CNOT gate between qubit 1 (control) and qubit 2 (target) belonging to each their logical state is given by the operation $\mathcal{H}_{2}(CZ_{12})\mathcal{H}_2$ where $\mathcal{H}_i$ is a Hadamard transform on the $i$'th qubit and $CZ_{ij}$ is a CZ-gate on qubits $i$ and $j$. The first Hadamard on the target qubit, however, results in information about the logical state being leaked to the environment if the subsequent CZ gate fails. This can be circumvented by first entangling qubit 2 (control) with an auxiliary qubit (target) as incorporated in \figref{fig:CNOTgate}. This entanglement with the auxiliary qubit can be achieved with an effective CNOT gate since the auxiliary qubit will not leak information about the logical state in case of a failed CZ-gate. An effective CNOT can then be performed between qubit 1 (control) and the auxiliary qubit (target) and due to the entanglement with qubit 2, the auxiliary qubit will not leak information about the logical state if the CZ gate fails. In that case, the three qubits are simply measured and the logical states can be recovered with single qubit rotations.  If the gate succeeds, qubit 2 is measured in the rotated basis and based on the measurement result, single qubit rotations are applied to the auxiliary qubit and qubit 1. This amounts to an effective CNOT between qubit 1 (control) and 2 (target) together with a state transfer of qubit 2 to the auxiliary qubit.  

We numerically simulate a logical CNOT gate assuming all measurements and single qubit rotations are deterministic and that the associated errors of these are negligible compared to errors due to e.g. limited coherence time of the qubits. Failed gates decrease the encoding level and we require final re-encoding of the logical qubits at the end of the logical gate operation to have a successful logical gate \cite{footnote}. To simulate final re-encoding, we have assumed that the qubits that have been removed from the encoding are recycled at the end of the logical gate operation and grown into a GHZ state, which is then fused with the qubit state as described above using an auxiliary qubit and a heralded CZ gate. 

The optimal performance of the logical CNOT gate for a given success probability of the heralded CZ-gate is found through numerical optimization. In the optimizations, we vary the encoding level $n$ and calculate the mean total error probability for each $n$. This error probability consists of both the probability of non-heralded errors, e.g. due to finite coherence time, and the probability that the logical encoding is lost because of too many failed attempts. While increasing $n$ decreases the probability of losing the logical encoding, it also decreases the effective coherence time of the logical state. From the simulations, we then find the encoding resulting in the smallest total error probability of the logical CNOT operation. We study two scenarios for implementing a logical CNOT gate. First, we describe a simple, local implementation where all qubits are trapped at the same node. Next, we describe a non-local implementation where the logical qubits are contained at different nodes in a network. Here, remote entanglement is used to teleport gate operations between the nodes resulting in a non-local logical CNOT gate for quantum networks. In both situations, we assume that a heralded CZ-gate can be performed between any two qubits at the same node.   

\begin{figure} [h]
\centering
\subfloat {\label{fig:figure2a}\includegraphics[width=0.45\textwidth]{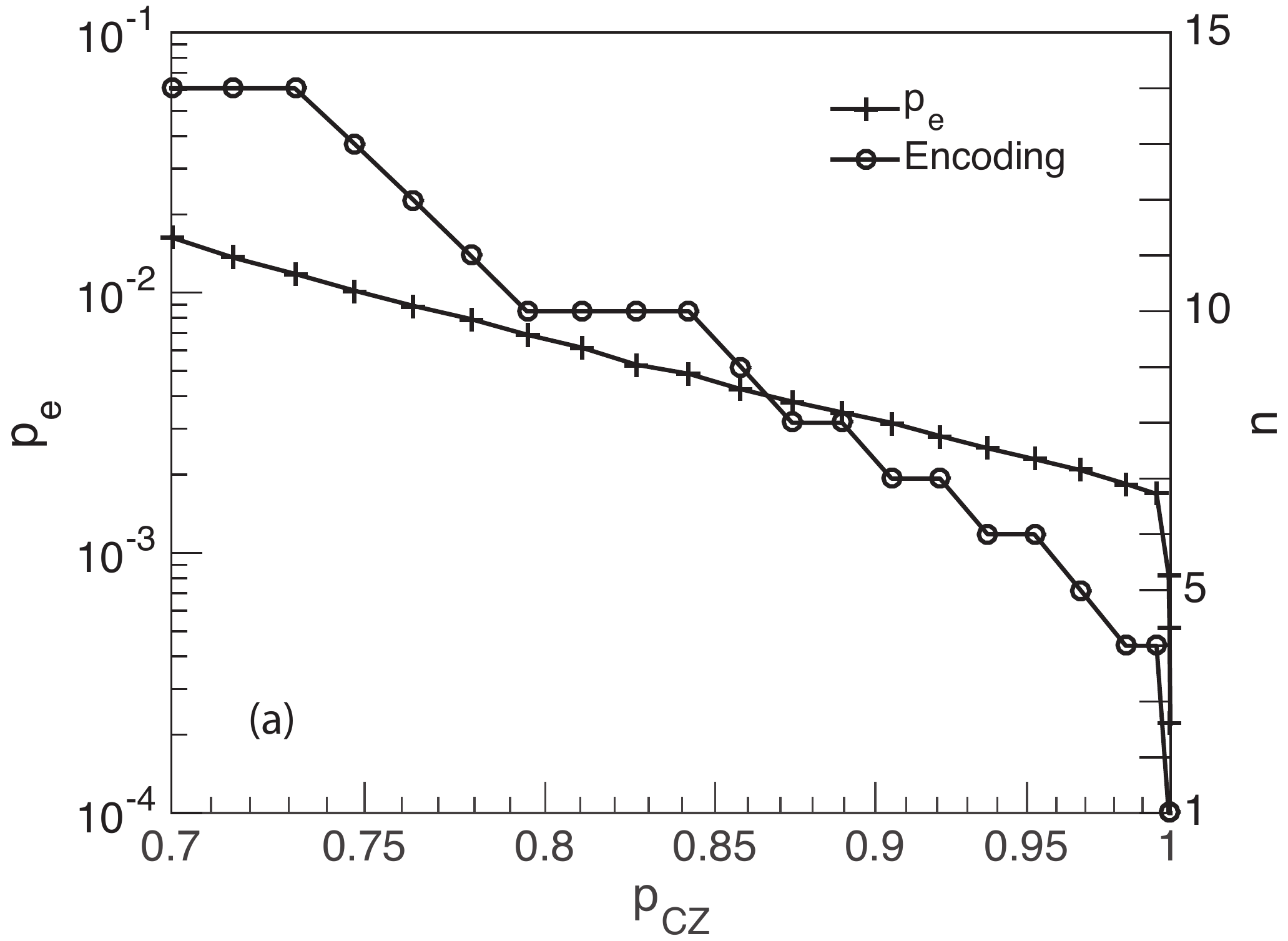}} \\
\subfloat {\label{fig:figure2b}\includegraphics[width=0.45\textwidth]{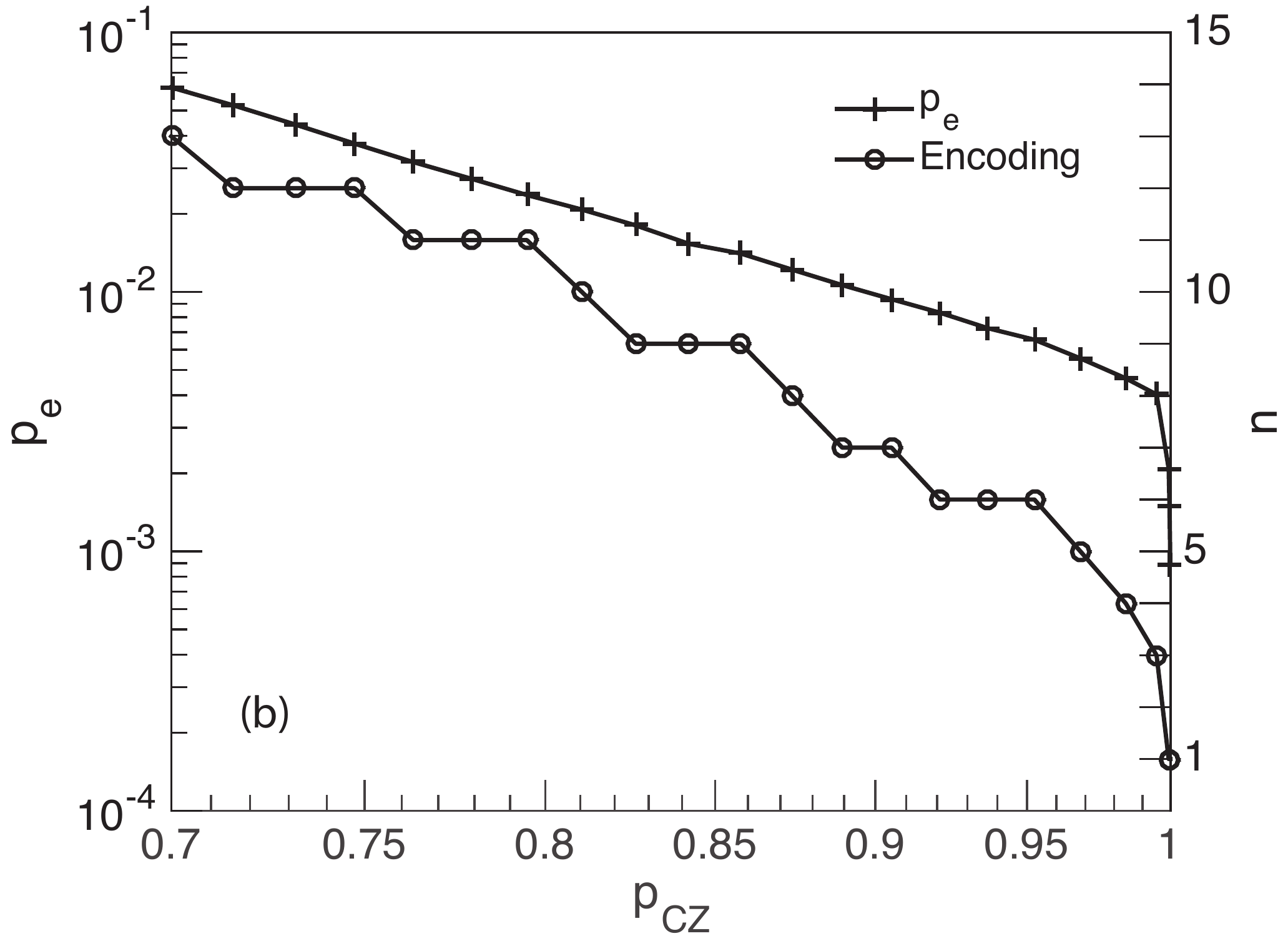}} 
\caption{Total error probability, $p_e$ (left axis) and encoding level, $n$ (right axis) of a (a) local and (b) non-local logical CNOT gate as a function of the success probability, $p_{\text{CZ}}$ of the heralded CZ-gate. We have assumed that all qubits have a coherence time of 1 s and that the error of the heralded CZ-gate is $10^{-4}$. The time of the CZ-gate is assumed to be 10 $\mu$s. Furthermore, we have assumed that the time necessary to make a Bell pair between two nodes (neighboring cavities) is 160 $\mu$s (10 $\mu$s) for the non-local implementation (see \figref{fig:figure3}). Each point in the plots is obtained from averaging over $5\cdot10^{4}$ Monte Carlo simulations where the gates are picked to succeed or fail at random according to $p_{\text{CZ}}$. For non-encoded states ($n=1$), a direct CNOT-gate is performed between the qubits without the use of a resource state.}
\label{fig:figure2}
\end{figure}

In order to simulate the logical CNOT, we assume that the coherence time of the physical qubits is 1 s. Although this coherence time might be challenging in some systems, it has been achieved both with trapped ions~\cite{langer} and in solid-state emitters~\cite{maurer1}. The gate time and error of the heralded CZ-gate are assumed to be 10 $\mu$s and $10^{-4}$ motivated by a realization with Rb atoms~\cite{borregaard1}. Note that fast and efficient atomic state readout can be achieved through the cavity field~\cite{rempe2}. The optimal performance and encoding level found from the simulations of a local CNOT gate are shown in \figref{fig:figure2a}.   
The error in \figref{fig:figure2a} is determined by the combination of a finite coherence time of the atoms and the gate time of the heralded CZ gate together with the error of the heralded CZ gate. The steep increase in the error probability shown in \figref{fig:figure2a} for near-deterministic CZ-gates is where no encoding is used and the failure probability of the gate directly adds to the error probability. However, as encoding is employed to correct for the failed attempts, the error only increases by a factor of $\sim10$ as the success probability of the CZ gate decreases from 99.5\% to 70\%. Even with a modest success probability of 75\% for the heralded gates, an error probability of $1\%$ can be attained. This demonstrates how the encoding can enable high fidelity, deterministic gates from heralded gates.   

\begin{figure} [h]
\centering
\includegraphics[width=0.35\textwidth]{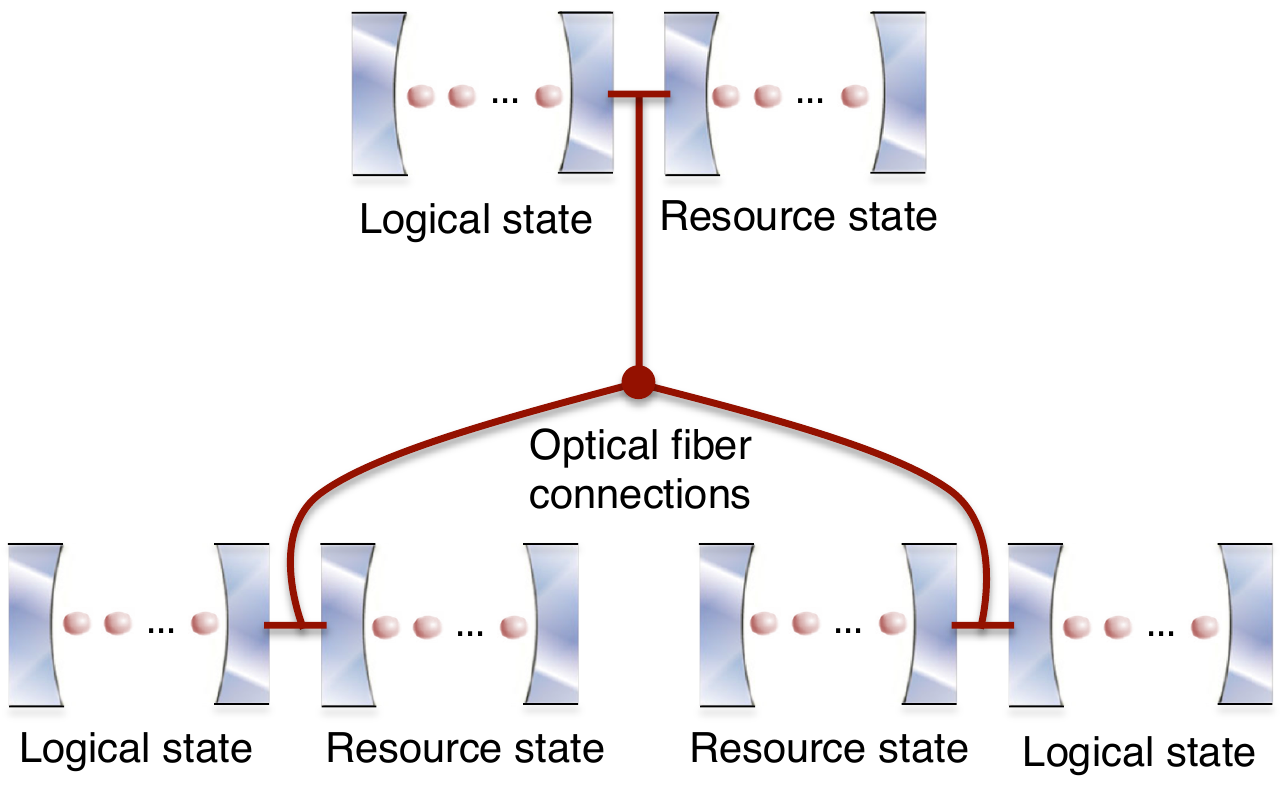}
\caption{Structure of the minimalistic quantum network with optical cavities considered for the simulation of the non-local CNOT gate. Each node consists of two cavities - one containing the encoded qubit state and one containing a resource state for logical operations. The nodes are assumed to be connected through optical fibers.}
\label{fig:figure3}
\end{figure}

We now describe the non-local implementation of the logical CNOT gate, which exploits the efficient interface between photons and atomic or atom-like qubits provided by a cavity. If a high fidelity Bell pair can be generated between two nodes, a non-local gate between the nodes can be obtained by teleporting the local gates through the Bell pair as described in Refs.~\cite{eisert1,anders3}. There exist many proposals for non-local entanglement generation~\cite{huelga,duan1,monroe1} but, in general, two-photon detection schemes~\cite{duan1,monroe1} obtain higher fidelity than one-photon detection schemes~\cite{huelga} and these have therefore been considered in great detail for quantum communication~\cite{borregaard2,sangouard1,sangouard2}. Inefficient photo detection, however, is detrimental to the rate in this case. Instead we consider the scheme of Refs.~\cite{cirac2,cirac1}. Here atomic detection, which in general is much more efficient, is employed instead of photo detection. Consequently, the rate of entanglement generation can be increased substantially while maintaining a high fidelity. To simulate a non-local version of the logical CNOT gate, we assume that such a scheme is used to generate high fidelity ($F\sim1$) entanglement between cavities in a minimalistic setup shown in \figref{fig:figure3}. The logical qubit and the resource state are assumed to be held in two different cavities that are spatially close to each other while optical fibers are used as channels between different nodes in the network.
We have numerically optimized the distributed CNOT gate and found the optimal encoding level for a given success probability of the heralded gate. 

The result of the optimization is shown in \figref{fig:figure2b}. As before, we have assumed a single qubit coherence time of 1 s and a heralded error of the CZ gate of $10^{-4}$. The gate time was assumed to be 10 $\mu$s while the Bell state preparation time was assumed to be 160 $\mu$s corresponding to internodal distances around 10 km assuming that fiber losses at telecom wavelength limit the rate. The entanglement generation time between cavities within the same node was assumed to be 10 $\mu$s. The heralded error of a Bell state was assumed to be  $10^{-4}$ and we have neglected any measurement error. Compared to \figref{fig:figure2a}, the error in \figref{fig:figure2b} is enhanced due to the relatively long preparation time of the Bell states. Furthermore, the scaling of the error with the success probability is less favorable due to the extra CZ-gates necessary to teleport the gate operation through the Bell pairs. Nonetheless, the error only increases by a factor $\sim15$ as the failure probability decreases from 99.5\% to 70\% in this minimalistic setup. An encoded error rate of 1\% can be obtained with a success probability of $\sim89\%$ for the heralded gates.  The combination of heralded schemes ensuring high-fidelity operations in cavity setups with logical encoding is thus a promising approach to realize high-fidelity operations in an extended quantum network.  

In conclusion, we have shown how heralded gate operations can be combined with logical encoding from LOQC to make efficient non-local operations in a quantum network. As opposed to the alternative method of using probabilistic gates for one way quantum computation our approach circumvents the generation of large graph states. In particular, it was argued in Ref.~\cite{raussendorf} that in order to simulate a quantum circuit of logical depth $d$ involving $n$ qubtis, an upper bound on the size of the graph state is on the order of $n^3d$. The approach considered here thus gives a more efficient approach towards distributed quantum computing~\cite{eisert1,beals} based on probabilistic gates, where separated resources are combined in order to perform a quantum algorithm. Given the difficulty in scaling up systems locally, the possibility of combining spatial separated resources might be an important step towards realizing realistic quantum computation. Non-local quantum gates are also naturally applicable in delegated quantum computing~\cite{broadbent,fisher} and secure multiparty function evaluation~\cite{benor,dupuis} where quantum operations are performed on encrypted data in order to ensure the privacy of the involved parties. Other applications of quantum networks such as quantum enhanced metrology~\cite{komar1}, anonymous quantum communication~\cite{brassard1} and cryptographic conferencing~\cite{chen1} use non-local GHZ states between the nodes as resources. Such non-local GHZ states could be obtained using the deterministic non-local CNOT gate described here to fuse smaller GHZ state together in a repeater like setup. 

\begin{acknowledgements}
We gratefully acknowledge the support from NSF, CUA, AFOSR, MURI, ARL, The Vannevar Bush Faculty Fellowship and the European Research Council under the European Union's Seventh Framework Programme (FP/2007-2013) through ERC Grant QIOS (Grant No. 306576). J.B. acknowledge support from the Carlsberg Foundation. 
\end{acknowledgements}

\end{document}